\documentclass[journal]{IEEEtran}
\usepackage{amsmath}
\usepackage{graphicx} 
\usepackage{makecell} 
\usepackage{graphicx} 
\usepackage{float} 

\usepackage{titlesec}

\titlespacing*{\subsubsection}{0pt}{0.8em}{0.4em}
\titlespacing*{\paragraph}{0pt}{0.6em}{0.3em}

\ifCLASSINFOpdf
\else
\fi
\begin{document}
%
\title{ECGXtract: Deep Learning-based ECG Feature
Extraction for Automated CVD Diagnosis}
%
%
%

\author{%
    Youssif~Abuzied,~Hassan~Abd\textendash Eltawab,~Abdelrhman~Gaber,~and~Tamer~ElBatt%
    \thanks{All authors are with the Department of Computer Science and Engineering, 
    The American University in Cairo (AUC), New Cairo 11835, Egypt 
    (e-mail: \{youssif.abuzied, hassan.abdeltawab, gaberabdo68, tamer.elbatt\}@aucegypt.edu).}%
    \thanks{This publication was developed as part of Afretec Network which is managed by Carnegie Mellon University Africa and receives financial support from the Mastercard Foundation. The views expressed are solely those of the authors and do not necessarily reflect those of Carnegie Mellon University Africa or the Mastercard Foundation.}%
}

\maketitle

\begin{abstract}
This paper presents ECGXtract, a deep learning–based approach for interpretable ECG feature extraction, addressing the limitations of traditional signal processing and black-box machine learning methods. In particular, we develop convolutional neural network models capable of extracting both temporal and morphological features with strong correlations to a clinically validated ground truth. 
Initially, each model is trained to extract a single feature, ensuring precise and interpretable outputs. A series of experiments is then carried out to evaluate the proposed method across multiple setups, including global versus lead-specific features, different sampling frequencies, and comparisons with other approaches such as ECGdeli. Our findings show that ECGXtract achieves robust performance across most features with a mean correlation score of 0.80 with the ground truth for global features, with lead II consistently providing the best results. For lead-specific features, ECGXtract acheives a mean correlation score of 0.822. Moreover, ECGXtract achieves superior results to the state-of-the-art open-source ECGdeli as it got a higher correlation score with the ground truth in 90\% of the features. Furthermore, we explore the feasibility of extracting multiple features simultaneously utilizing a single model. Semantic grouping is proved to be effective for global features, while large-scale grouping and lead-specific multi-output models show notable performance drops. These results highlight the potential of structured grouping strategies to balance the computational efficiency vs. model accuracy, paving the way for more scalable and clinically interpretable ECG feature extraction systems in limited resource settings.\end{abstract}


\section{Introduction}
Cardiovascular diseases (CVDs) remain a leading global health challenge, causing about 17.3 million deaths annually—nearly 37\% of worldwide mortality—a number projected to reach 23.6 million by 2030 \cite{plawiak2020deep}. Beyond their high mortality, CVDs impose substantial economic burdens due to long-term treatment costs, straining healthcare systems and reducing patients’ quality of life. Early diagnosis and timely treatment are therefore critically important \cite{chen2019smart}. The economic impact also extends to indirect costs such as productivity loss, particularly affecting low- and middle-income countries \cite{gheorghe2018economic}.

The electrocardiogram (ECG) is a vital diagnostic tool that records cardiac electrical activity over time. It captures electrical impulses generated by muscle contractions essential for proper heart function. The ECG waveform consists of the P wave, QRS complex, and T wave—representing atrial depolarization, ventricular depolarization, and ventricular repolarization, respectively—allowing clinicians to detect abnormalities \cite{wasilewski2011introduction}. The ECG signal comprises repetitive cardiac cycles, as shown in Figure 1. However, due to its complexity and potential irregularities, manual interpretation is time-consuming and error-prone, limiting scalability. Thus, computer-aided diagnostic tools are increasingly needed to ensure accurate and efficient ECG analysis \cite{abagaro2024automated}.

A major approach in automated CVD diagnosis involves developing machine learning algorithms to identify cardiac abnormalities in ECG recordings \cite{jafari2023deep}. These models improve diagnostic accuracy by learning patterns from large labeled ECG datasets. Techniques such as convolutional neural networks (CNNs), recurrent neural networks (RNNs), and ensemble methods have achieved strong results in classifying arrhythmias and other conditions from raw or preprocessed ECG data \cite{kiranyaz2021real, yildirim2020accurate}.

\begin{figure}[!t]
\centering
\includegraphics[width=\columnwidth]{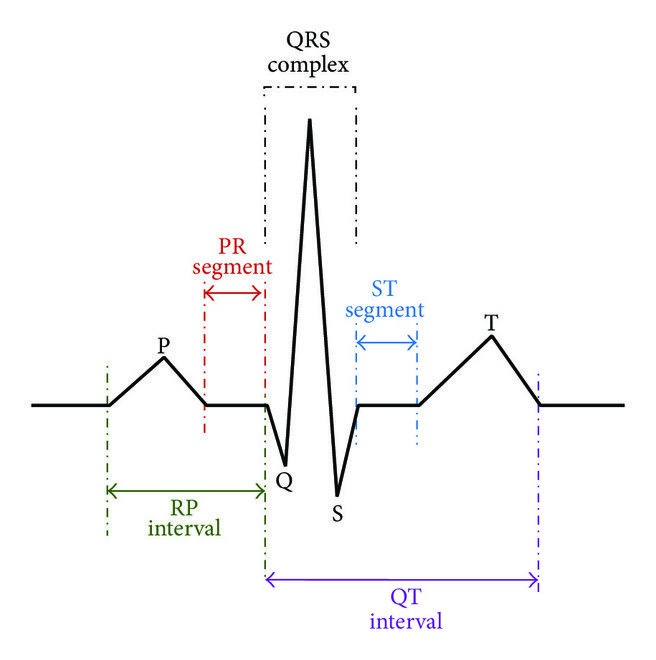}
\caption{A sample ECG waveform.}
\label{fig:cardiac_cycle}
\end{figure}

An alternative strategy focuses on extracting ECG features to help physicians detect trends and abnormalities more effectively \cite{jafari2023deep}. As noted in \cite{arenas2020svm}, ECG features fall into morphological (shape-based) and temporal (time-based) categories. Morphological features include QRS duration, ST-segment deviation, and P-wave amplitude, while temporal features include RR intervals, heart rate variability (HRV), and QT intervals \cite{litfl_ecg_diagnosis}. Automated extraction using signal processing or deep learning methods enhances diagnostic accuracy and supports real-time remote monitoring.

ECG morphological and temporal features are crucial for clinicians to interpret cardiac characteristics and diagnose CVDs \cite{zhao2005wavelet}. Extensive research has applied machine learning algorithms to distinguish healthy and unhealthy patients, where feature extraction remains a key determinant of performance \cite{singh2023trends}. Existing work generally falls into three directions. The first employs machine learning techniques such as principal component analysis (PCA) and linear discriminant analysis (LDA) to derive features, though these are often not clinically interpretable \cite{gupta2021novel}. The second leverages digital signal processing (DSP) techniques—such as wavelet transform (WT) and short-term Fourier transform (STFT)—to extract clinically relevant elements like R peaks and QRS complexes \cite{gupta2021review}, but struggles with complex parameters like RP intervals. The third involves proprietary commercial algorithms capable of extracting hundreds of features \cite{strodtthoff2023ptbxlplus}, though their high cost limits accessibility, especially in low-resource settings.

This work extends the authors' previous study in \cite{abuzied2025automated}. The contributions are threefold. First, we present a simple yet effective deep learning model for accurately extracting both temporal and morphological ECG features—the first, to our knowledge, to achieve interpretable feature extraction via machine learning. Second, unlike prior methods limited to specific features, our model is inherently extensible to additional ECG parameters. Third, ECGXtract achieves high computational efficiency, requiring significantly less training and inference time than comparable methods. We perform extensive preprocessing on raw ECG signals before feeding them into our deep neural network, which then produces the target features efficiently and accurately.

The rest of this paper is organized as follows. Section 2 reviews related work and its limitations. Section 3 introduces the proposed methodology and its underlying rationale. Section 4 presents experimental results and performance analysis. Section 5 concludes with key findings and directions for future research.
\section{Related Work}
This section examines the research domain most pertinent to our study, specifically machine learning-based ECG feature extraction, to underscore the contributions of prior work and delineate the limitations that motivate the present investigation.

In their systematic review of the usage of deeplearning models in ECG data, \cite{hong2020opportunities} highlighted that current deep learning approaches for ECG feature extraction automatically capture morphological and temporal signal characteristics, enabling high-accuracy classification and biometric recognition. Most methods rely on short-term or outdated datasets, limiting generalizability. Complex model architectures often function as black boxes, reducing interpretability of the extracted features, which is critical in medical applications. High computational requirements hinder deployment on portable or real-time devices. Additionally, integration with traditional expert-designed features is limited, and models frequently struggle with imbalanced disease labels or single-modality ECG data. These challenges constrain the reliability and applicability of current deep learning–based ECG feature extraction methods.

Machine learning–based approaches for extracting features from ECG signals have garnered significant scholarly interest.  As demonstrated by  \cite{kuila2020feature} , one ECG-based biometric recognition framework employs stages of data preprocessing, P-QRS-T wave detection, and subsequent feature extraction. The system extracts morphological and temporal attributes—namely time intervals, amplitudes, slopes, and angles—and applies classifiers such as artificial neural networks (ANNs), support vector machines (SVMs), and K-nearest neighbors (KNNs). This methodology yields high recognition accuracy; however, it is encumbered by substantial computational burden due to the reliance on wavelet transforms, time-domain analyses, and normalization procedures, rendering it less suitable for real-time implementation. Moreover, the limited feature set—comprising up to 72 features—may constrain adaptability for future applications that require richer or higher-dimensional representations. 

In an effort to automate ECG feature extraction, \cite{montenegro2022evaluation} proposed a one-dimensional convolutional neural network (CNN) to streamline the process. While this approach improves efficiency, it suffers from a significant limitation: the features extracted by the CNN lack direct correspondence to the morphological and temporal characteristics of the ECG signal—such as P waves, QRS complexes, and T waves—that are essential for clinical interpretation and diagnosis by cardiologists.
For the same purpose of automating ECG features extraction but with more focus on interpretability, authors in \cite{verma2022development} developed CNN and LSTM models for time-series ECG classification. They found that the models primarily relied on the QRS complex for predictions, and interpretability techniques such as Grad-CAM and Permutation Feature Importance provided both local and global explanations. Limitations included class imbalance, limited patient diversity, and challenges in explaining dense ECG signals, while PDP and SHAP proved less effective due to computational or interpretability constraints.

Zhang et al. \cite{zhang2024fusion} proposed a feature extraction methodology that converts one-dimensional ECG signals into two-dimensional images, integrating multi-lead recordings into dual-channel inputs. Their CNN architecture incorporates a multi-scale pyramid module with dilated convolutions to capture contextual information at multiple scales. However, this method is limited by high computational requirements, dependency on large datasets, and the reduced interpretability of the abstract features generated by the network.

To enhance classification of cardiovascular diseases, Lou et al. \cite{lou2023latent} applied Empirical Mode Decomposition (EMD) to break down ECG signals into Intrinsic Mode Functions (IMFs), which were subsequently reconstructed as two-dimensional feature maps for processing with a 2D CNN. Despite improving classification performance, this method is constrained by high computational demands and the reduced interpretability of the extracted features. In a similar approach, Hasan and Bhattacharjee \cite{hasan2019deep} proposed a method for multi-class cardiovascular disease classification using ECG signals processed through Empirical Mode Decomposition (EMD). The first three Intrinsic Mode Functions (IMFs) were combined to form a modified ECG signal, which was then input to a one-dimensional convolutional neural network (CNN). While this approach improves feature learning and denoising, it is computationally intensive and its reliance on EMD may limit interpretability of the extracted features. The method achieved high classification accuracies of 97.70\%, 99.71\%, and 98.24\% on three public databases.

Perhaps the closest to our work is the study by Dhara et al. \cite{dhara2024adaptive}, which employs Deep Convolutional Neural Networks (DCNNs) for ECG feature extraction, emphasizing interpretable morphological and temporal characteristics. In this approach, ECG signals are first pre-processed to reduce noise and standardize the data, after which they are fed into a DCNN to extract hierarchical features from multiple signal components, including the QRS complexes, P waves, T waves, and R-R intervals. The extracted deep features are then fused and classified using an Enhanced Radial Basis Function (ERBF) network for the diagnosis of various heart diseases. While this method demonstrates notable effectiveness and accurate feature extraction, it is limited by its high computational complexity, which complicates real-time implementation. Additionally, the model exhibits sensitivity to noise, potentially affecting feature reliability, and its dependence on specific datasets raises concerns regarding its applicability and generalizability to broader clinical scenarios.

\section{Methodology}
The proposed methodology involves the design, development and training of a deep neural network model using a large publicly available dataset hosting ECG signals along with their corresponding labeled features. This section provides a comprehensive description of the methodological framework, encompassing the dataset utilized, the proposed architecture of the model, and the training strategy employed to optimize the model.

\subsection{The PTB-XL+ Dataset}
Identifying a publicly available dataset hosting ECG signals along with their corresponding extracted features is a crucial step for the proposed methodology. Among the most comprehensive resources of this kind is the PTB-XL+ dataset \cite{strodtthoff2023ptbxlplus}. This dataset is derived from the original PTB-XL ECG collection \cite{wagner2020ptbxl}, and comprises 21,799 clinical 12-lead ECG recordings obtained from 18,869 patients, each with a duration of 10 seconds. Furthermore, the ECG signals are available in two sampling frequencies, namely 500 Hz and 100 Hz. The PTB-XL+ dataset extends the original PTB-XL by incorporating features extracted through three distinct approaches: two based on IP-protected commercial products \cite{macfarlane2005unig, gehealthcare2019marquette} and one employing an open-source DSP-based technique \cite{pilia2021ecgdeli}. These extracted features are standardized into a unified set of well-defined and interpretable descriptors. Additionally, the dataset specifies which features are extracted by which extraction method and provides correlation analysis between the features obtained via different methods. The dataset categorizes the features into two main types:
\begin{itemize}
\item Lead-based features: Features that depend on a specific lead.
\item Global features: Features that are independent of any single lead.
\end{itemize}

The PTB-XL+ dataset is selected for training and evaluating the proposed model due to its large sample size and the extensive variety of interpretable ECG features it offers. The features provided by \cite{macfarlane2005unig} were utilized as the ground truth for both training and testing. This is attributed to the fact that this method yielded the highest number of features and demonstrates strong correlations with the other two feature extraction methods documented in the dataset.

\subsection{Machine Learning for ECG Feature Extraction}
In this subsection, we provide an overview of the machine learning models that have been applied to ECG feature extraction. We first review prior approaches, ranging from traditional statistical models to more advanced deep learning methods, and then present the architecture of the proposed CNN-based framework. 

\subsubsection{Prior Machine Learning Models for ECG Feature Extraction}

In pursuit of the stated objective, multiple machine learning models were evaluated. First, linear regression or logistic regression are tried as they are capable of capturing simple correlations in ECG features (for example, a direct relationship between heart rate and a specific cardiac disease). Additionally, support vector machines (SVMs) have long been used for ECG classification due to their strong generalization ability on small or high-dimensional datasets. For instance, one study shows that an SVM achieved superior ECG beat classification performance compared to k-NN and radial basis function networks, clearly confirming the superiority of the SVM approach over traditional classifiers \cite{melgani2008svm}. 

For modeling temporal structure, recurrent networks like LSTM (Long Short-Term Memory) are well‐suited to ECG data. LSTMs are specifically designed to find patterns in time. Because ECG is a sequential time series of beats, the ability of LSTMs to remember characteristic fragments of the signal (e.g. previous PQRST complexes) is crucial. In practice, this means an LSTM can retain information about earlier waveform shapes and learn long-range dependencies in the heartbeat sequence, which often leads to improved arrhythmia or feature prediction performance on ECG data \cite{klosowski2020lstm}. 

Deep CNN (Convolutional Neural Network) models have also proven very effective for ECG feature extraction. CNNs are hierarchical networks inspired by the visual cortex; they excel at learning spatial filters that detect local patterns, even in one-dimensional time series. In ECG analysis, CNNs automatically learn salient waveform features (such as the QRS complex shape or ST-segment variations) directly from the raw signal, without manual feature design. As noted in the literature, CNNs have been widely used for time-series classification (including arrhythmia detection) because they detect important features automatically without the need for human intervention \cite{botros2022cnnsvm}.


\subsubsection{The Proposed Model Architecture}
\begin{figure*}[!t]
    \centering
    \includegraphics[width=\textwidth]{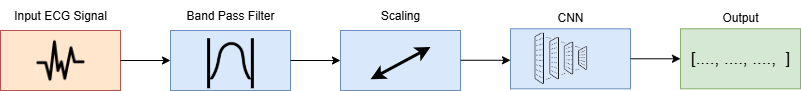}
    \caption{Main stages in ECGXtract.}
    \label{fig:SystemStages}
\end{figure*}

\begin{figure*}[!t]
    \centering
    \includegraphics[width=\textwidth]{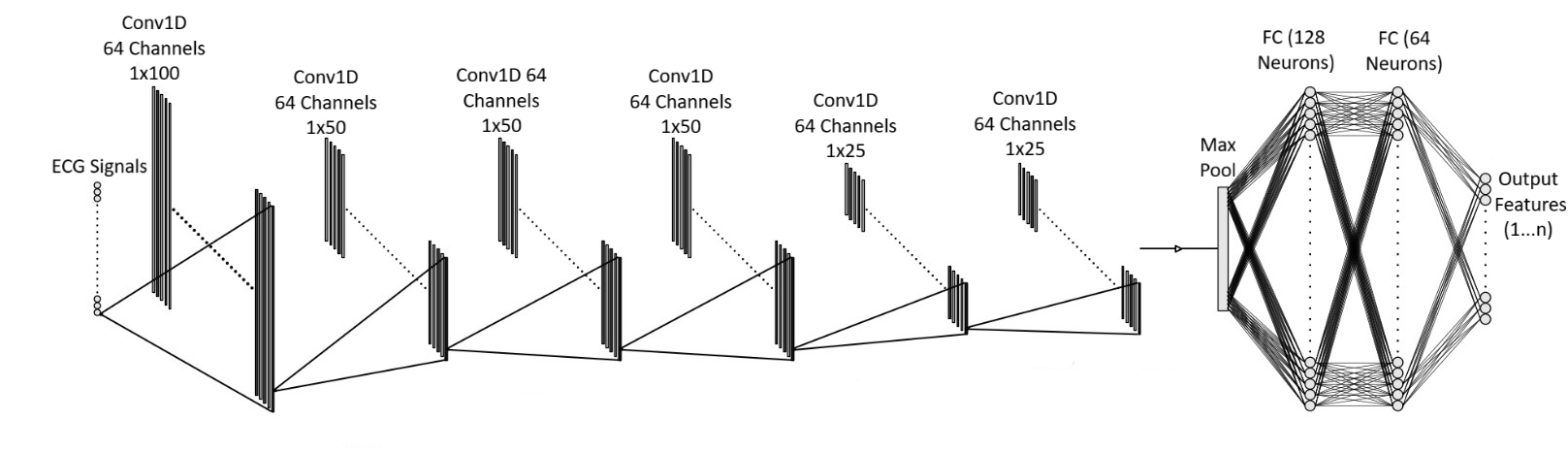}
    \caption{The proposed CNN model architecture for ECG feature extraction.}
    \label{fig:ModelArchitecture}
\end{figure*}

We first experimented with linear regression, SVMs, and LSTMs. However, all of them resulted in a modest correlation score of approximately 0.60. This relatively limited performance motivated the investigation of using CNNs. In comparison to the aforementioned models, CNNs demonstrated superior performance in terms of both correlation with the ground truth and training efficiency. The CNN-based model attained an average correlation of 0.78, as will be detailed in Section IV. The main components and stages in ECGXtract are shown in Figure \ref{fig:SystemStages} Figure \ref{fig:ModelArchitecture} also highlights the detailed architecture of the selected CNN model.


The design process began with a simple configuration consisting of two convolutional layers followed by a fully connected layer, with the model depth incrementally increased until no additional performance improvement was observed. Extensive hyperparameter optimization was performed, involving variations in kernel sizes, activation functions, and other CNN parameters. The objective was to construct an effective and computationally efficient model capable of supporting ECG feature extraction in low-resource environments with constrained processing capabilities.

\subsection{Training Procedure}
This section outlines the primary challenges encountered and the key design decisions implemented during model training to address these issues and enhance overall performance.

This section presents the techniques employed for ECG signals pre-processing, a critical component in ensuring effective model convergence. A prominent challenge posed by the adopted ECG dataset stems from its inherently noisy nature, which significantly impedes or even prevents the model from converging. To mitigate this, a carefully selected band-pass filter was applied. According to professional guidelines, ECG frequencies predominantly lie between 0.5 Hz and 150 Hz, and filtering outside of this range effectively removes both baseline wanders and high-frequency noise \cite{salimi2023ecg}. Other scholars further suggest that a practical and widely adopted passband for non-diagnostic ECG applications is 0.5 Hz to 40 Hz, which balances noise reduction while preserving essential waveform features, yielding higher-quality signals and fewer non-interpretable traces \cite{ricciardi2016impact}. Consequently, the implementation of a 0.5–40 Hz band-pass filter reduced the adverse impact of noise in the dataset while preserving pertinent ECG attributes.

The second challenge concerned the scaling of target features, which was found to have a direct influence on convergence speed. To accelerate both training and evaluation, min–max scaling was applied to the target features prior to training. Following model inference, the generated features were transformed back to their original scale for evaluation against the corresponding ground truth values.

While the proposed architecture is capable of extracting multiple features simultaneously, preliminary observations suggest that the number of features processed per model can  influence performance. Training multiple features together may introduce interference, potentially leading to less optimal feature representations, whereas different features may require distinct representational structures to generalize effectively. The optimal strategy for determining the number of features to be extracted per model will be systematically investigated in the experiments section.

\section{Experimental Evaluation}
This section presents the primary evaluation metric employed, describes the experiments conducted to assess the performance and limitations of the proposed ECG feature extraction methodology, and reports as well as discusses the main results. A total of six experiments are included. All experiments are conducted on Google Colab notebooks, a computational platform with 15 GB of GPU RAM and 12 GB of system RAM. Unless otherwise noted, all experiments are conducted at a sampling rate of 500 Hz. Experiment~1 evaluates the ability of the proposed model to extract global features when training a separate model for each feature individually. Experiment~2 investigates whether multiple global features can be extracted simultaneously using a single model. Experiment~3 shifts the focus to lead-specific features, with models trained to predict each feature one at a time, while Experiment~4 explores the feasibility of extracting multiple lead-specific features concurrently within one model. Experiment~5 examines the effect of varying the sampling frequency (500 Hz vs. 100 Hz) on model performance. Finally, Experiment~6 benchmarks the proposed ECGXtract approach against the ECGdeli method introduced in \cite{pilia2021ecgdeli}.

\subsection{The Correlation Performance Metric}
The performance of the proposed method was assessed using the \textit{Pearson Correlation Coefficient} (\( PCC \)), a widely adopted statistical measure for quantifying linear correlation between two variables, that is, features in our case \cite{sedgwick2012pearson}. It is mathematically defined as:
\[
pcc = \textstyle \scriptsize \frac{n \sum_{i=1}^{n} y_{\text{test},i} \hat{y}_i - \left( \sum_{i=1}^{n} y_{\text{test},i} \right) \left( \sum_{i=1}^{n} \hat{y}_i \right)}
{\sqrt{ \left[ n \sum_{i=1}^{n} y_{\text{test},i}^2 - \left( \sum_{i=1}^{n} y_{\text{test},i} \right)^2 \right] 
\left[ n \sum_{i=1}^{n} \hat{y}_i^2 - \left( \sum_{i=1}^{n} \hat{y}_i \right)^2 \right] }}
\]

where,
\begin{itemize}
    \item \( y_{\text{test},i} \) and \( \hat{y}_i \) represent the ground truth and predicted values, respectively.  
    \item \( n \) is the number of data points.  
    \item \( \bar{y}_{\text{test}} \) and \( \bar{\hat{y}} \) are the mean values of the ground truth and predictions, respectively.  
\end{itemize}

\subsection{Extraction of Global Features}
This subsection includes all the experiments pertaining to the extraction of global  features, either by training a separate model for each feature or by training a model for a group of features. 
\subsubsection{Experiment 1: The First Step - Extracting One Global ECG Feature}

\paragraph{Experiment Setup} 
In this experiment, we aim to prove the effectiveness of our method to extract one global feature for each trained model. Given that exhaustively testing all possible lead combinations would be both computationally expensive and time-consuming, the selection of configurations is guided by insights from prior studies in the literature.

The first configuration utilizes all 12 ECG leads, serving as a benchmark for our experiment. The second configuration is inspired by \cite{issa2023heartbeat}, which proposed a high-accuracy heartbeat classification method relying solely on lead II; accordingly, only lead II is used in this setup. The third configuration is based on the findings of \cite{lai2021optimal}, where a subset of four leads (II, aVR, V1, V4) is shown to enhance the generalizability of deep learning models for ECG abnormality classification. Thus, this combination is adopted as the third setup. Finally, to support the goal of cost-efficient ECG feature extraction, a fourth configuration consisting of six leads (I, II, III, aVF, aVR, aVL) is employed, as these can be recorded using only four electrodes, which is the most common setup in ECG devices.
\paragraph{Results}

Table~\ref{tab2} summarizes the average training and inference durations, computed over 1000 ECG records, for each of the four lead configurations. As shown, using a single lead or a small subset of leads substantially reduces both training and inference times compared to using all 12 leads.

\begin{table}[htbp]
\caption{Training and Inference Times (per 1000 ECG records) for the Four Setups}
\label{tab2}
\centering
\begin{tabular}{|c|c|c|c|c|}
\hline
\textbf{Metric} & \textbf{\textit{\makecell{All\\Leads}}} & \textbf{\textit{II}} & \textbf{\textit{\makecell{II, aVR,\\V1, V4}}} & \textbf{\textit{\makecell{I, II, III,\\aVF, aVR, aVL}}} \\
\hline
\textbf{Training Time (min.)} & 45.6 & 15.4 & 23.2 & 30.1 \\
\hline
\textbf{Inference Time (s)} & 4.15 & 1.39 & 2.14 & 2.89 \\
\hline
\end{tabular}
\end{table}

Table~\ref{tab1} reports the Pearson correlation coefficients (PCC) between the extracted features and their corresponding ground truth values across the four configurations.

\begin{table}[!t]
\caption{Experiment I PCC Results}
\label{tab1}
\centering
\begin{tabular}{|c|c|c|c|c|}
\hline
\textbf{Global Feature} & \textbf{\textit{\makecell{All\\Leads}}} & \textbf{\textit{II}} & \textbf{\textit{\makecell{II, aVR,\\V1, V4}}} & \textbf{\textit{\makecell{I, II, III,\\aVF, aVR, aVL}}} \\
\hline
\textbf{T\_Off} & 0.855 & 0.870 & 0.871 & 0.857 \\
\hline
\textbf{QRS\_Off} & 0.830 & 0.849 & 0.848 & 0.829 \\
\hline
\textbf{HR\_Ventr} & 0.870 & 0.885 & 0.886 & 0.872 \\
\hline
\textbf{QRS\_On} & 0.847 & 0.850 & 0.849 & 0.846 \\
\hline
\textbf{RR\_Mean} & 0.810 & 0.816 & 0.815 & 0.812 \\
\hline
\textbf{QT\_IntFramingham} & 0.802 & 0.796 & 0.797 & 0.801 \\
\hline
\textbf{QT\_IntFridericia} & 0.786 & 0.803 & 0.802 & 0.785 \\
\hline
\textbf{QT\_Int} & 0.790 & 0.786 & 0.787 & 0.791 \\
\hline
\textbf{QT\_IntBazett} & 0.778 & 0.782 & 0.784 & 0.779 \\
\hline
\textbf{T\_On} & 0.772 & 0.775 & 0.773 & 0.773 \\
\hline
\textbf{QT\_IntCorr} & 0.765 & 0.769 & 0.768 & 0.766 \\
\hline
\textbf{P\_Off} & 0.758 & 0.763 & 0.761 & 0.759 \\
\hline
\textbf{QRS\_Dur} & 0.752 & 0.756 & 0.752 & 0.753 \\
\hline
\textbf{P\_On} & 0.745 & 0.749 & 0.748 & 0.746 \\
\hline
\textbf{PR\_Int} & 0.738 & 0.742 & 0.740 & 0.739 \\
\hline
\textbf{P\_AxisFront} & 0.732 & 0.735 & 0.736 & 0.733 \\
\hline
\end{tabular}
\end{table}

Overall, the four evaluated configurations demonstrated comparable performance across all global features. The setups using lead~II alone and the combination of leads (II, aVR, V1, V4) consistently achieved the highest correlation scores. In particular, the lead~II-only configuration produced the best results for 11 out of the 16 global features. Including all 12 leads did not yield any performance improvement, indicating that most features can be effectively extracted from lead~II or a small subset of leads.

These results align with previous studies, where lead~II is commonly used as the primary reference in ECG analysis~\cite{issa2023heartbeat, yildirim2020accurate}. Moreover, the reduced training and inference times associated with lead~II make it especially suitable for real-time and resource-constrained applications. Retraining the model for additional features using only lead~II is therefore expected to be significantly more efficient. Consequently, employing lead~II alone—or in combination with a limited set of leads—offers an optimal trade-off between performance and computational cost.

\subsubsection{Experiment 2: Extracting Multiple Global Features Using a Single Model}

\paragraph{Experiment Setup} 

In the previous experiment, each model was trained to extract a single global feature, which, while offering the highest accuracy since the model is tuned specifically for that feature, is computationally inefficient because each feature must be processed individually. In this experiment, we extend the setup to investigate whether multiple global features can be extracted concurrently using a single model, trading some potential loss in accuracy for significantly improved computational efficiency. The main motivation is to reduce the number of model runs while preserving accuracy as much as possible, thereby improving both efficiency and interpretability.

We explored two strategies for grouping features within a model. The first is \textit{Random Grouping}, where features are paired arbitrarily, without consideration of their semantic relationships. The second is \textit{Semantic Grouping}, which is inspired by the idea that semantically related outputs can often be learned jointly with minimal or no loss of accuracy, whereas unrelated features may interfere with each other (“feature interference”) \cite{hackernoon2023multioutput}. In this setting, the 16 global features from Experiment 1 are grouped into 8 semantically coherent pairs (i.e., two features per group) based on the feature categories in the original dataset.
\paragraph{Results}

Table~\ref{tab7} outlines the grouping schemes used for both the random and semantic approaches. 

\begin{table}[!t]
\caption{Comparison of Random and Semantic Feature Groupings}
\label{tab7}
\centering
\resizebox{\columnwidth}{!}{%
\begin{tabular}{|c|cc|cc|}
\hline
\textbf{Group} & \multicolumn{2}{c|}{\textbf{Random Groups}} & \multicolumn{2}{c|}{\textbf{Semantic Groups}} \\
\hline
- & \textbf{\makecell{Feature 1}} & \textbf{\makecell{Feature 2}} & \textbf{\makecell{Feature 1}} & \textbf{\makecell{Feature 2}} \\
\hline
1 & \makecell{RR\_Mean} & \makecell{QT\_Int\\Fridericia} & \makecell{QT\_Int\\Framingham} & \makecell{QT\_Int\\Bazett} \\
\hline
2 & \makecell{QRS\_Off} & \makecell{QT\_Int\\Corr} & \makecell{QT\_Int} & \makecell{QT\_Int\\Corr} \\
\hline
3 & \makecell{T\_On} & \makecell{P\_Axis\\Front} & \makecell{QRS\_On} & \makecell{QRS\_Off} \\
\hline
4 & \makecell{QT\_Int\\Bazett} & \makecell{QRS\_On} & \makecell{P\_On} & \makecell{P\_Off} \\
\hline
5 & \makecell{QT\_Int\\Framingham} & \makecell{P\_Off} & \makecell{T\_On} & \makecell{T\_Off} \\
\hline
6 & \makecell{PR\_Int} & \makecell{QT\_Int} & \makecell{RR\_Mean} & \makecell{HR\_Ventr} \\
\hline
7 & \makecell{HR\_Ventr} & \makecell{P\_On} & \makecell{QRS\_Dur} & \makecell{PR\_Int} \\
\hline
8 & \makecell{QRS\_Dur} & \makecell{T\_Off} & \makecell{P\_Axis\\Front} & \makecell{QT\_Int\\Fridericia} \\
\hline
\end{tabular}%
}
\end{table}

Table~\ref{tab8} then compares the feature-wise performance across the original, random, and semantic grouping strategies. The results indicate that semantic grouping generally maintained accuracy more effectively than random grouping, with several features even showing small improvements.

\begin{table}[!t]
\caption{Comparison of Feature Scores for Semantic and Random Grouping}
\label{tab8}
\centering
\resizebox{\columnwidth}{!}{%
\begin{tabular}{|c|c|c|c|}
\hline
\textbf{\makecell{Feature}} & 
\textbf{\makecell{Single Feature \\ Extraction PCC}} & 
\textbf{\makecell{Semantic \\ Grouping PCC}} & 
\textbf{\makecell{Random \\ Grouping PCC}} \\
\hline
\makecell{T\_Off} & 0.870 & 0.872 & 0.809 \\
\hline
\makecell{QRS\_Off} & 0.849 & 0.820 & 0.808 \\
\hline
\makecell{HR\_Ventr} & 0.885 & 0.888 & 0.809 \\
\hline
\makecell{QRS\_On} & 0.850 & 0.810 & 0.806 \\
\hline
\makecell{RR\_Mean} & 0.816 & 0.818 & 0.746 \\
\hline
\makecell{QT\_Int\\Framingham} & 0.796 & 0.799 & 0.727 \\
\hline
\makecell{QT\_Int\\Fridericia} & 0.803 & 0.764 & 0.767 \\
\hline
\makecell{QT\_Int} & 0.786 & 0.747 & 0.713 \\
\hline
\makecell{QT\_Int\\Bazett} & 0.782 & 0.800 & 0.724 \\
\hline
\makecell{T\_On} & 0.775 & 0.773 & 0.703 \\
\hline
\makecell{QT\_Int\\Corr} & 0.769 & 0.729 & 0.699 \\
\hline
\makecell{P\_Off} & 0.763 & 0.761 & 0.722 \\
\hline
\makecell{QRS\_Dur} & 0.756 & 0.718 & 0.689 \\
\hline
\end{tabular}%
}
\end{table}

To further examine this trend, we extended the grouping strategy to include models predicting more than two features at once, forming flexible sets based on physiological similarity (e.g., QT interval, QRS complex). This approach aimed to enhance computational efficiency while minimizing accuracy loss. Table~\ref{tab9} summarizes the final feature groups.

\begin{table}[!t]
\caption{Feature Groups and their Corresponding ECG Features}
\label{tab9}
\centering
\begin{tabular}{|l|p{5cm}|}
\hline
\textbf{Group} & \textbf{Features} \\
\hline
QT Interval–Related & QT\_Int, QT\_IntCorr, QT\_IntBazett, \newline QT\_IntFramingham, QT\_IntFridericia \\
\hline
QRS Complex Timing & QRS\_On, QRS\_Off, QRS\_Dur \\
\hline
P Wave Timing & P\_On, P\_Off, P\_AxisFront \\
\hline
T Wave Timing & T\_On, T\_Off \\
\hline
Heart Rate \& Interval & RR\_Mean, HR\_Ventr, PR\_Int \\
\hline
\end{tabular}
\end{table}

Table~\ref{tab10} compares the grouped model performance to single-feature extraction. Although certain features showed moderate degradation, most retained acceptable accuracy, indicating that several global features can be reliably extracted within one model when guided by semantic relationships.

\begin{table}[!t]
\caption{Comparison of Grouped and Original Feature Values}
\label{tab10}
\centering
\resizebox{\columnwidth}{!}{%
\begin{tabular}{|c|c|c|c|c|}
\hline
\textbf{\makecell{Feature}} & 
\textbf{\makecell{Group}} & 
\textbf{\makecell{Grouped \\ Features}} & 
\textbf{\makecell{Original \\ Values}} & 
\textbf{\makecell{Difference}} \\
\hline
\makecell{T\_Off} & \makecell{Heart Rate \\ \& Interval} & 0.869 & 0.870 & -0.001 \\
\hline
\makecell{QRS\_Off} & \makecell{QT Interval\\–Related} & 0.832 & 0.849 & -0.017 \\
\hline
\makecell{HR\_Ventr} & \makecell{QRS Complex \\ Timing} & 0.855 & 0.885 & -0.030 \\
\hline
\makecell{QRS\_On} & \makecell{QT Interval\\–Related} & 0.807 & 0.850 & -0.043 \\
\hline
\makecell{RR\_Mean} & \makecell{T Wave \\ Timing} & 0.814 & 0.816 & -0.002 \\
\hline
\makecell{QT\_Int\\Framingham} & \makecell{P Wave \\ Timing} & 0.768 & 0.796 & -0.028 \\
\hline
\makecell{QT\_Int\\Fridericia} & \makecell{QT Interval\\–Related} & 0.800 & 0.803 & -0.003 \\
\hline
\makecell{QT\_Int} & \makecell{QRS Complex \\ Timing} & 0.749 & 0.786 & -0.037 \\
\hline
\makecell{QT\_Int\\Bazett} & \makecell{QT Interval\\–Related} & 0.769 & 0.782 & -0.013 \\
\hline
\makecell{T\_On} & \makecell{P Wave \\ Timing} & 0.771 & 0.775 & -0.004 \\
\hline
\makecell{QT\_Int\\Corr} & \makecell{Heart Rate \\ \& Interval} & 0.755 & 0.769 & -0.014 \\
\hline
\makecell{P\_Off} & \makecell{QT Interval\\–Related} & 0.721 & 0.763 & -0.042 \\
\hline
\makecell{QRS\_Dur} & \makecell{Heart Rate \\ \& Interval} & 0.757 & 0.756 & +0.001 \\
\hline
\makecell{P\_On} & \makecell{P Wave \\ Timing} & 0.747 & 0.749 & -0.002 \\
\hline
\makecell{PR\_Int} & \makecell{QRS Complex \\ Timing} & 0.696 & 0.742 & -0.046 \\
\hline
\makecell{P\_Axis\\Front} & \makecell{T Wave \\ Timing} & 0.737 & 0.735 & +0.002 \\
\hline
\end{tabular}%
}
\end{table}

Experiment~2 provides valuable insight into the feasibility of extracting multiple global features using shared models. Initially, comparing random and semantic pairings revealed that random grouping led to substantial degradation, with correlations often decreasing by more than 0.05. In contrast, semantic grouping largely preserved performance and occasionally improved it—for instance, features such as QT\_IntBazett and HR\_Ventr maintained or exceeded their single-feature scores. This confirms that physiologically related outputs benefit from shared representational structures, making semantic grouping a more robust and efficient strategy, consistent with findings in~\cite{hackernoon2023multioutput}.

Extending this idea to larger semantic clusters (e.g., all QT interval–related descriptors) yielded moderate success but introduced increasing interference as group size grew. Features like PR\_Int and QRS\_On experienced declines exceeding 0.04 in PCC, suggesting that overly broad groupings reduce specificity. Still, several features—such as QRS\_Dur and P\_AxisFront—retained near-original accuracy, indicating that careful, clinically meaningful grouping can balance efficiency and precision. Overall, the results highlight a clear trade-off: semantic grouping reduces computational cost and model count but excessive aggregation harms accuracy. A practical solution is to treat these grouped models as shared bases and fine-tune them for specific features, following approaches like Task-Adaptive Parameter Sharing (TAPS)~\cite{wallingford2022task}, which achieves efficient multi-task specialization by adapting only a small subset of parameters for each task.

\subsection{Extraction of Lead-Specific Features}

Lead-specific features differ from global features in that they depend directly on the lead by which the signal is recorded. This makes their extraction more challenging, as performance can vary across leads. To study this, we designed two complementary experiments: the first focuses on extracting one lead-specific feature per model, while the second explores combining multiple lead-specific features within a single model.

\subsubsection{Experiment 3: Extracting Individual Lead-Specific Features}

\paragraph{Experiment Setup}

In this experiment, our objective is to evaluate the model's ability to extract lead-specific features when trained on a single feature at a time. Since each feature exists across all 12 leads, we processed each lead independently, computed the correlation for every lead, and then derived the average correlation across leads. However, average values can obscure important differences, as some leads may be more informative than others. To address this, we also report the variance across leads and highlight the lead with the maximum PCC correlation score for each feature. 

\paragraph{Results}
Table~\ref{tab3} summarizes the results, showing strong average correlations and modest variations in performance across leads.

\begin{table}[!t]
\caption{Lead-Specific Feature Extraction Results}
\label{tab3}
\begin{center}
\footnotesize
\begin{tabular}{|c|c|c|c|}
\hline
\textbf{Lead-Specific Feature} & \textbf{\makecell{PCC}} & \textbf{Variance} & \textbf{\makecell{Lead with max \\ performance}} \\
\hline
QRS\_AmpPP & 0.892 & 0.0021 & II \\
\hline
R\_Dur & 0.885 & 0.0032 & V5 \\
\hline
R\_Amp & 0.869 & 0.0018 & II \\
\hline
QRS\_Area & 0.857 & 0.0037 & aVF \\
\hline
S\_Amp & 0.851 & 0.0015 & V3 \\
\hline
T\_Area & 0.842 & 0.0029 & II \\
\hline
S\_Dur & 0.843 & 0.0012 & V6 \\
\hline
Q\_Dur & 0.822 & 0.0035 & II \\
\hline
Q\_Amp & 0.810 & 0.0009 & aVL \\
\hline
Rp\_Dur & 0.794 & 0.0024 & II \\
\hline
ST\_Amp & 0.779 & 0.0017 & V1 \\
\hline
P\_Area & 0.761 & 0.0030 & III \\
\hline
Sp\_Dur & 0.759 & 0.0011 & V4 \\
\hline
Rp\_Amp & 0.747 & 0.0026 & II \\
\hline
\end{tabular}
\end{center}
\end{table}
Experiment~3 demonstrates that the proposed model effectively extracts lead-specific features with correlation levels comparable to global ones. The low variance values indicate consistent performance across leads, with deviations typically within $\pm0.05$. Among all leads, Lead~II achieved the highest correlation in 6 out of the 14 features, reaffirming its reliability for ECG feature extraction. Notably, no other lead achieved the top score for more than one feature, underscoring the distinctive importance of Lead~II in representing both temporal and morphological ECG characteristics.

\subsubsection{Experiment 4: Extracting Multiple Lead-Specific Features Using a Single Model}

\paragraph{Experiment Setup}
While Experiment 3 focused on one feature per model, in this experiment we investigate whether multiple lead-specific features can be reliably extracted using a single model. Each lead-specific feature appears across all 12 leads, resulting in 12 feature instances. To explore the trade-off between efficiency and performance, we grouped features in increasing sizes—pairs, triplets, sets of four, sets of six, and finally all twelve features—training a single model for each group.

During training, the model receives as input the ECG signals corresponding to all instances in the group (e.g., for a group of three lead-specific features, the signals for all three features across all leads are fed into the model simultaneously). The model is then expected to output the predicted values for all features in the group concurrently.

\begin{figure}[!t]
\centering
\includegraphics[width=\columnwidth]{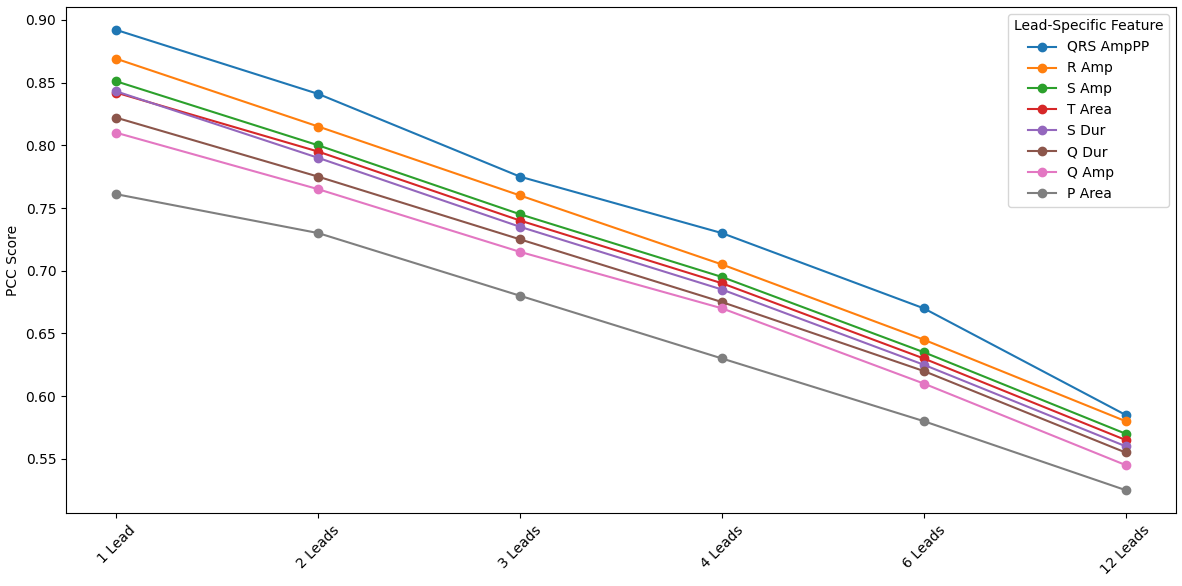}
\caption{PCC performance across different grouping configurations for lead-specific features.}
\label{fig:PCC}
\end{figure}

For evaluation, we compute a single representative value per feature as follows: for a given feature, we first obtain the predicted score for each instance in the group, then average the scores within that group. Finally, we average these group-level scores across all groups containing the feature to obtain one value per feature. This setup allows us to analyze how grouping affects predictive accuracy and whether feature interactions are beneficial or detrimental.
\paragraph{Results}
The PCC scores for various grouping configurations, shown in Figure \ref{fig:PCC}, indicate that moderate grouping maintains competitive performance, while excessive grouping causes clear drops in correlation—emphasizing the need to balance efficiency with feature-level accuracy.

Experiment 4 explores whether multiple lead-specific features can be jointly extracted within a single model. The results show a consistent decline in performance as the number of grouped features increases. Combining two or three features preserves strong correlations, with occasional slight gains, but grouping six or all twelve features results in over 0.20 correlation loss compared to individual training. This indicates that lead-specific features need specialized representations and may interfere when processed together in a shared architecture. Unlike global features that benefit from semantic grouping, lead-specific ones appear less compatible with broad multi-output modeling because each lead captures unique, non-aligning signal patterns.

Overall, while small-scale grouping of lead-specific features can improve efficiency, training individual or minimally grouped models remains the most dependable strategy for accurate extraction. Future research should systematically explore which lead combinations can be grouped effectively without compromising performance.

\subsection{Experiment 5: The Role of the ECG Sampling Frequency}
\paragraph{Experiment Setup}
In this experiment, we investigate the impact of the ECG signal sampling frequency on performance. Specifically, we compare the correlation results obtained when running our proposed model at two sampling rates: the 500~Hz used in the dataset and common in hospital-grade ECG equipment for clinical use vs. a downsampled 100~Hz version. The primary objective is to determine whether changing the sampling frequency leads to performance degradation and, if so, to quantify its extent. Since the previous two experiments were conducted at 500~Hz, we adopt the best-performing configuration from Experiment~1 for global features and the average correlation score from Experiment~3 for lead-specific features as baselines. For the comparison, we selected 4 global features and 4 lead-specific features. The features were selected based on their diversity in performance in Experiments 1 and 3, ensuring that the chosen set included features that differed from each other in terms of correlation and behavior. This approach allows the comparison to capture a representative range of feature characteristics rather than redundant or similar features. For global features, we reuse the optimal setup from Experiment~1 with 100~Hz signals, while for lead-specific features, we replicate the procedure from Experiment~2 using the 100~Hz data.

\begin{figure}[!t]
    \centering
    \includegraphics[width=\columnwidth]{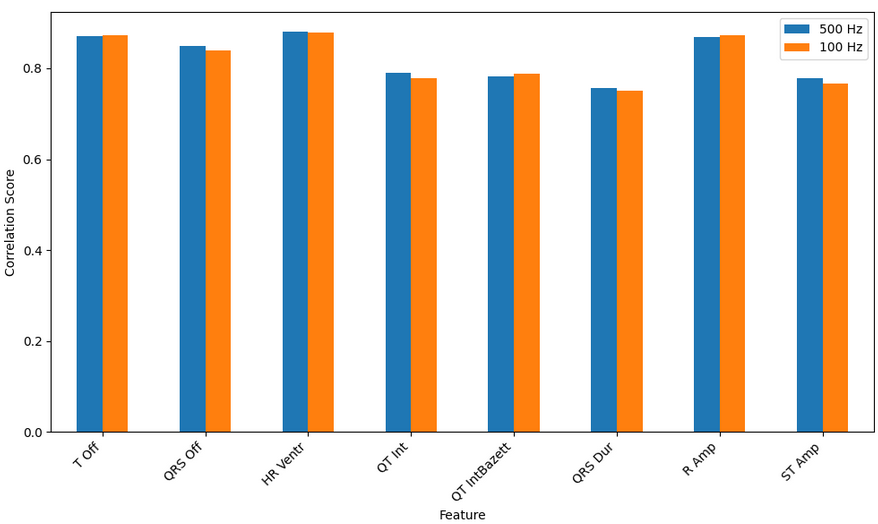}
    \caption{Correlations of model trained on 500 Hz signals and models trained on 100 Hz signals.}
    \label{fig:Correl}
\end{figure}

\paragraph{Results}
The bar chart in Figure \ref{fig:Correl} presents the changes in correlation scores across the two sampling rates for a subset of the 30 features. Table~\ref{tab5} reports the average training and inference times (measured over 1000 ECG records) for both frequencies.

\begin{table}[!t]
\caption{Comparison of Training and Inference Times for the Two Signal Frequencies}
\label{tab5}
\begin{center}
\begin{tabular}{|c|c|c|}
\hline
\textbf{Metric} & \textbf{\makecell{500 Hz}} & \textbf{\makecell{100 Hz}} \\
\hline
\textbf{Training Time (min.)} & 16.96 & 9.23 \\
\hline
\textbf{Inference Time (s)} & 1.47 & 0.83 \\
\hline
\end{tabular}
\end{center}
\end{table}

The results from Experiment~5 indicate that both sampling frequencies produce very similar correlation scores, suggesting that the lower 100~Hz sampling rate can be used without compromising model performance. This consistent performance is observed for both global and lead-specific features. Furthermore, utilizing the 100~Hz frequency not only maintains high correlation across all features but also reduces computational complexity, enabling faster training and inference. These observations are consistent with previous studies, which demonstrate that models based on convolutional neural networks can achieve high detection accuracy at 100~Hz while benefiting from reduced model complexity~\cite{habib2020sampling}.

\subsection{Experiment 6: Performance Comparison: ECGXtract against  The Baseline, ECGdeli \cite{pilia2021ecgdeli}}
\paragraph{Experiment Setup}
In this experiment, we aim to compare the performance of ECGXtract against a well-known method proposed earlier in the literature, coined ECGdeli \cite{pilia2021ecgdeli}. As illustrated in earlier sections of the paper, ECGdeli is an open-source toolbox, based on digital signal processing, for ECG delineation and feature extraction. It detects fiducial points using wavelet-based methods, derives morphological and temporal measurements, and outputs clinically interpretable results. The framework supports multi-lead analysis, enabling extraction of more than one hundred features \cite{pilia2021ecgdeli}. We had an interest in comparing the performance of ECGXtract to this method as it is accessible by anyone for using it and also requires minimal computation time. Extracted features from this method could be found in the datasets provided by authors in \cite{strodtthoff2023ptbxlplus}. Since the Uni-G method serves as our ground truth, we computed the Pearson correlation coefficient (PCC) for the features shared between Uni-G and ECGdeli, and then compared these values with the PCC between our extracted features and Uni-G. The top 10 highly correlated features between ECGdeli and Uni-G were selected for comparison. We used the 500 Hz ECG signals as the inputs to our models. For the extraction of the global features, we used only lead~II signals. For the lead-specific features, we calculated the average of 12 features in both methods. In this experiment, we relied on scores obtained in the one-feature-per-model setup.

\begin{figure}[!t]
    \centering
    \includegraphics[width=\columnwidth]{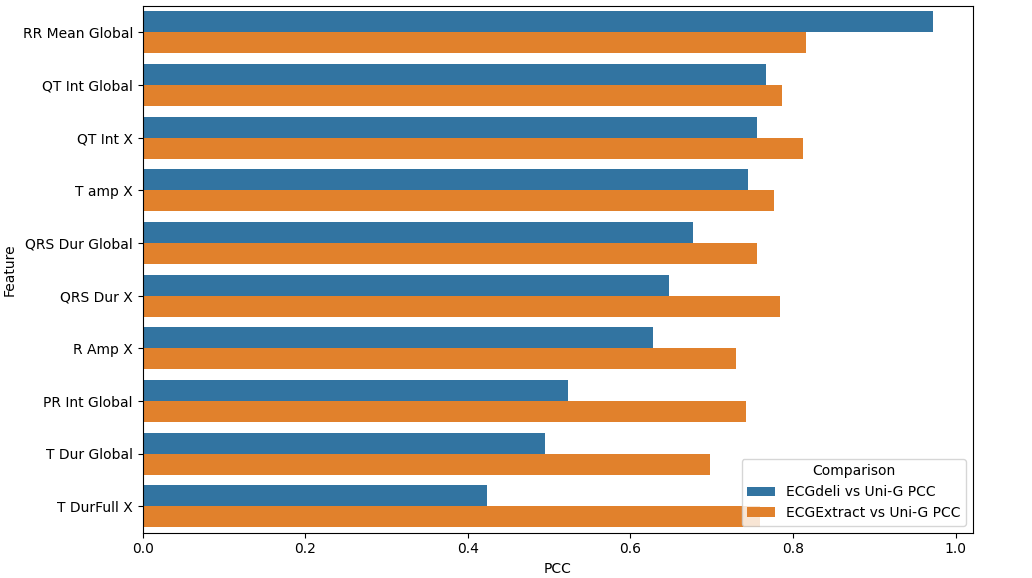}
    \caption{Correlation scores of features extracted using ECGDeli vs correlation scores of features extracted using ECGXtract.}
    \label{fig:Bar}
\end{figure}

\paragraph{Results}
Results of this experiment are listed in Figure \ref{fig:Bar}. Lead-specific features have suffix ``\_x'' and global features have the suffix ``\_global''.

The results from experiment ~6 clearly shows that ECGXtract yielded higher correlation results with the Uni-G approach than ECGdeli. In 9 out of the 10 listed features, ECGXtract got a higher correlation with Uni-G method. The lone exception is RR\_Mean\_Global, where ECGdeli remains substantially closer to Uni-G — a predictable outcome since RR estimation depends on precise R-peak detection, an area where deterministic QRS detectors are highly optimized. For the other features (intervals, durations and several amplitude/area measures) the learned model more closely reproduces Uni-G, likely because it can learn dataset-specific corrections and nonlinear relationships that deterministic delineation does not capture. That said, higher concordance with Uni-G is not proof of absolute correctness: some of the gains may reflect alignment between our training labels, preprocessing choices and Uni-G’s definitions rather than universally better measurements. To clarify causes we recommend per-lead analyses (to separate aggregation/lead-choice effects), scatter and Bland–Altman plots for the most discrepant features, and statistical tests (bootstrapped 95\% CIs and paired tests) to confirm significance. For future attempts, researchers might consider a combined approach: augment a deterministic R-peak detector for RR-related tasks (or incorporate an auxiliary R-peak head during training), and use the learned model for other features where it matches Uni-G better. Checking the ECGs that have the largest errors can also reveal patterns and help improve the model.

\section{Conclusion}
In this work, we introduced and evaluated a deep learning–based model designed to extract clinically interpretable ECG features. Unlike many existing approaches that emphasize classification or rely on abstract representations, ECGXtract directly predicts temporal and morphological descriptors, offering improved transparency and clinical utility. Across extensive experiments, we demonstrated that the model consistently achieves strong correlations with ground truth features. Notably, Lead II emerged as a reliable input for both global and lead-specific features, while reducing the sampling frequency to 100 Hz did not compromise accuracy. Comparisons with ECGdeli further confirmed the competitiveness of the proposed method. We also investigated the practicality of training multi-output models: semantic grouping of global features preserved performance and even yielded improvements in certain cases, whereas larger or arbitrary groupings introduced interference. For lead-specific features, grouping beyond two or three outputs resulted in substantial accuracy loss, highlighting the need for more specialized models. Nevertheless, the problem is ripe, and this work lays the foundation for promising future directions. In particular, reliance on a single dataset raises concerns about generalizability, and computational demands may hinder real-time clinical deployment. Future research should therefore explore cross-dataset validation, adaptive grouping strategies, and lightweight architectures suitable for portable devices. By addressing these challenges, the proposed approach can advance toward more efficient, scalable, and clinically impactful ECG analysis systems.
\IEEEpeerreviewmaketitle

\end{document}